\documentclass[aps,prl,superscriptaddress,showpacs,preprint,onecolumn]{revtex4}
 \usepackage{amsmath}  
 \usepackage{makeidx}
 \usepackage{graphicx}
 \usepackage{amsfonts}
 \usepackage[ansinew]{inputenc}
 \usepackage[usenames,dvipsnames]{pstricks}
 \usepackage{subfigure}
 \usepackage{epstopdf}
\usepackage{epsfig}
 \usepackage{mathrsfs}

              \makeindex

\begin{document}
\title{Alpha Cluster Formation {\bf and Decay} in Quartetting Wave Function Approach}
\author{Chang Xu}
\email{cxu@nju.edu.cn}
\affiliation{School of Physics, Nanjing University, Nanjing 210093, China}
\author{G. R\"{o}pke}
\email{gerd.roepke@uni-rostock.de}
\affiliation{Institut f\"{u}r Physik, Universit\"{a}t Rostock, D-18051 Rostock, Germany}
\affiliation{National Research Nuclear University (MEPhI), 115409 Moscow, Russia}
\author{P. Schuck}
\email{schuck@ipno.in2p3.fr}
\affiliation{Institut de Physique Nucl\'{e}aire, Universit\'e Paris-Sud, IN2P3-CNRS, UMR 8608, F-91406, Orsay, France}
\affiliation{Laboratoire de Physique et Mod\'elisation des Milieux Condens\'es, CNRS- UMR 5493,
F-38042 Grenoble Cedex 9, France}
\author{Zhongzhou Ren}
\email{zren@nju.edu.cn}
\affiliation{School of Physics, Nanjing University, Nanjing 210093, China}
\author{Y. Funaki}
\affiliation{RIKEN Nishina Center, Wako 351-0198, Japan}
\author{H. Horiuchi}
\affiliation {Research Center for Nuclear Physics (RCNP), Osaka University, Osaka  567-0047, Japan}
\affiliation {International Institute for Advanced Studies, Kizugawa 619-0225,  Japan}
\author{A. Tohsaki}
\affiliation{Research Center for Nuclear Physics (RCNP), Osaka University, Osaka 567-0047, Japan}
\author{T. Yamada}
\affiliation{Laboratory of Physics, Kanto Gakuin University, Yokohama 236-8501, Japan}
\author{Bo Zhou}
\affiliation{Faculty of Science, Hokkaido University, Sapporo 060-0810, Japan}


\date{\today}

\begin{abstract}
We present a microscopic calculation of $\alpha$-cluster formation in heavy nuclei by using the quartetting wave function approach.
The interaction of the quartet with the core nucleus is taken in local density approximation. The $\alpha$-cluster formation is found to be particularly sensitive to the interplay of the mean field felt by the $\alpha$-cluster and the Pauli blocking as a consequence of antisymmetrization.  The striking feature of $\alpha$-cluster formation probability across the major shell closures of 82 protons and 126 neutrons is reproduced.  The shell (or subshell) effects on the $\alpha$-cluster formation in superheavy nuclei are also analyzed.

\end{abstract}


\maketitle

{\it Introduction.} Although the radioactive $\alpha$-decay is an important issue dating back to the early days of nuclear physics, the $\alpha$-cluster formation problem as a major challenge in $\alpha$-decay theory has still not been fully understood up to now \cite{Mang60}. The description of $\alpha$-cluster formation in heavy nuclei, in principle, involves a complicated many-body problem and is very difficult to handle technically \cite{Delion09,DLW}. Only a few microscopic calculations were carried out to estimate the $\alpha$-cluster formation probability in the typical nucleus $^{212}$Po ($\alpha$+$^{208}$Pb) \cite{Varga,Patial}. The fully microscopic treatment of $\alpha$-like correlations in heavy nuclei is still not feasible with present computer capabilities. This is in contrast to the situation in light nuclei where the $\alpha$-like correlations of selfconjugate nuclei have been investigated extensively using microscopic approaches \cite{THSR,RGM,GCM,FMD,AMD}. Systematics of $\alpha$-cluster formation probability $P_{\alpha}$ in heavy nuclei have been considered for a long time \cite{Brown,Hatsukawa}. Empirical values of $P_{\alpha}$ were extracted from measured data by using analytical formulas or semi-classical approximations \cite{Brown,Hatsukawa,Batchelder,Xu,Denisov,Buck}.  The main message is that there exists a striking change of $\alpha$-cluster formation probability across the closed shells, especially the neutron shell closure $N=126$ and proton one $Z=82$. This shell (or subshell) effect may also be crucial for the $\alpha$-decay of superheavy nuclei \cite{Hamilton}.

{\it Method.} In this work, $\alpha$-cluster formation in both heavy and superheavy nuclei is investigated by a quartetting (four-nucleon) wave function approach, which is inspired by the THSR wave function concept for light nuclei and has been successfully applied to the $\alpha$-decay of $^{212}$Po \cite{Po,Xu2016}. In this approach, the wave function of quartetting state is subdivided in a unique way in the center of mass (c.o.m.) part and the intrinsic part \cite{Po,Xu2016}.
The separation of the c.o.m. motion is a key to simplify the problem of $\alpha$-cluster formation. A coupled system of wave equations is obtained describing the c.o.m. motion and a similar equation for the intrinsic motion. To make the approach practicable, we use a local-density approximation for the lead core nucleus neglecting the derivative terms of the intrinsic wave function. The Schr\"odinger equation for c.o.m motion contains the kinetic part as well as the potential part which is approximated by an effective c.o.m. potential $ W({r})=W^{\rm intr}( r)+W^{\rm ext}( r)$ \cite{Po,Xu2016}. The intrinsic part $W^{\rm intr}( r)$ approaches for large $r$ the bound state energy of the free $\alpha$-particle: $E_\alpha^{(0)} = -28.3$ MeV. This binding energy will be reduced at short distance $r$  because of Pauli blocking effects. The external part $W^{\rm ext}(r)$ is determined by the mean-field interaction $V_\tau^{\rm mf}( r)$ including the strong nucleon-nucleon interaction as well as the Coulomb interaction. Using the two-potential approach \cite{Gurvitz1988}, the effective c.o.m. potential is separated into two parts at the separation point $r_{\rm sep}$. By solving the corresponding c.o.m. Schr\"odinger equation, the bound state wave function $\Phi(r)$ will be calculated. Here the $\alpha$-transition probability is given as product of the formation probability $P_{\alpha}$, the pre-exponential factor $\nu$, and the exponential factor ${\cal T}$ \cite{Xu2016}. The $\alpha$-cluster formation probability $P_{\alpha}$ is obtained by integrating the bound state wave function $\Phi(r)$ from the critical radius $r_c$
(corresponding to the critical density $n_B^{\rm Mott}$, see below) to infinity \cite{Po,Xu2016}:
\begin{eqnarray}
P_{\alpha}=\int_0^\infty  d^3r |\Phi(r)|^2 \Theta
\left[n_B^{\rm Mott}-n_B(r)\right]\,.
\end{eqnarray}
In the region $r > r_{\rm sep} $, the scattering state wave function $\chi(r)$ is obtained as a combination of regular and irregular
Coulomb functions \cite{Gurvitz1988,Xu2016}. The decay width is then calculated by using the values of $\Phi(r)$ and $\chi_k(r)$ at the
separation point $r_{\rm sep}$ \cite{Gurvitz1988,Xu2016}.

{\it Model parameters.} To avoid the problems of odd-nucleon blocking and angular momentum, only the favored transitions of even-even $\alpha$-emitters are considered. Firstly we focus on the most important case, \textit{i.e.} the $\alpha$-decay of polonium isotopes (Po $\rightarrow$ Pb+ $\alpha$). We {start} with the nucleon density in the lead core and determine the critical radius $r_c$ where the $\alpha$-like bound state is dissolved \cite{Po}: $n_B(r_c)=0.02917$ fm$^{-3}$. We use the following neutron and proton densities for the lead nucleus \cite{Tarbert2014}
\begin{eqnarray}
&& n_n( r)=\frac{N}{1343.62}/[1+e^{(r-6.7)/0.55}], \nonumber \\
&& n_p( r)=\frac{Z}{1303.76}/[1+e^{(r-6.68)/0.447}]
\end{eqnarray}
(in units of fm). To calculate the different isotopes of Po, we fix the proton number $Z=82$ and vary the neutron number $N$ only. The Pauli blocking term is determined by the baryon density $n_B$=$n_n$+$n_p$ with a fitted formula (in units of MeV, fm) $W^{\rm Pauli}(n_B)=4515.9\, n_B -100935\, n_B^2+1202538\, n_B^3$, which is valid in the density region $n_B \le 0.03$ fm$^{-3}$ with relative error below 1\% \cite{Po}. Both the nuclear and Coulomb potentials of the $\alpha$-core system are obtained from a double-folding model using the matter (and charge) densities of the core and the $\alpha$-particle. For the nuclear potential, the M3Y-type nucleon-nucleon interaction with a short-range repulsion part ($c$) and a long-range attraction part ($d$) are used, $v(s)=c \exp(-4s)/(4s)-d \exp(-2.5s)/(2.5s)$, in which $s$ denotes the nucleon-nucleon distance \cite{M3YReview}. In a first attempt, the parameters $c$ and $d$ are fitted to the experimental $\alpha$-decay energy ($Q_{\alpha}$) and half-life ($T_{1/2}$) for each polonium isotope \cite{Audi}. Below we discuss the systematics of these parameters which shows the predictive power of our approach.

\begin{figure}[thb]
\includegraphics[width=0.6\textwidth]{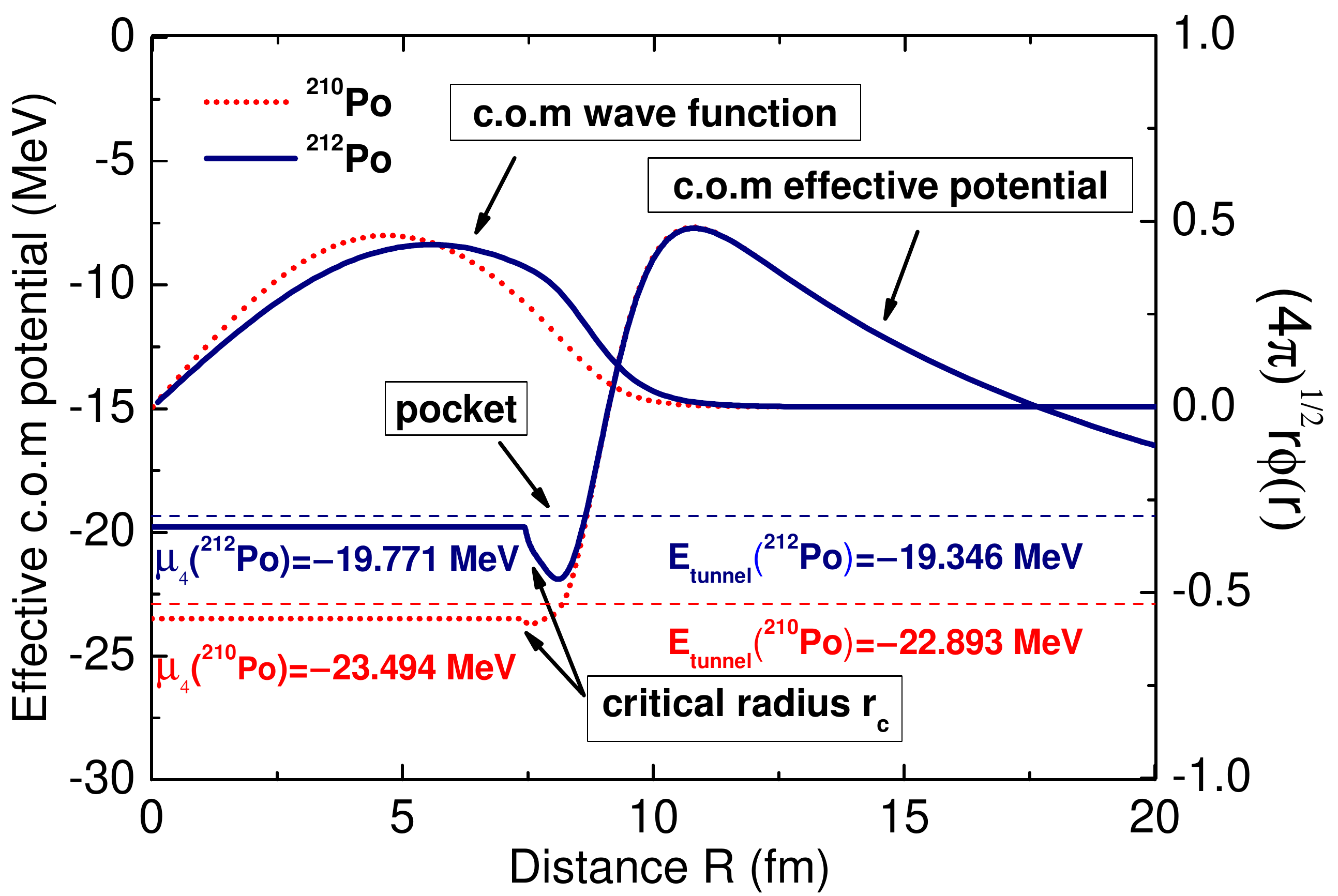}
\caption{(Color online) Comparison of the c.o.m. effective potentials, the c.o.m. wave functions, and the Fermi energies for two neighbouring $\alpha$-emitters $^{210}$Po and $^{212}$Po.}
 \label{Fig:1}
 \end{figure}

{\it Results.} In Fig.~\ref{Fig:1}, we compare the c.o.m effective potentials and the corresponding c.o.m motion wave functions for two neighbouring polonium isotopes $^{210}$Po and $^{212}$Po. Note that their empirical $\alpha$-cluster formation probability has the most significant change across the $N=126$ major shell. As shown in Fig.~\ref{Fig:1},
the c.o.m effective potentials of both $^{210}$Po and $^{212}$Po are dominated by the Coulomb repulsion for large
distances. At short distances, both the
attractive nuclear potential and repulsive Pauli blocking between
the $\alpha$-cluster and the lead core become relevant.  At a
critical radius $r_c$, the
$\alpha$-cluster is suddenly dissolved and the four nucleons added
to the core are implemented on top of the Fermi energy $\mu_4$
\cite{Po}. The critical radii are $r_c(^{210}{\rm Po})=7.432$ fm and $r_c(^{212}{\rm Po})=7.438$ fm, respectively. The bound state energy $E_{\rm tunnel}$ is above the Fermi energy $\mu_4$ for the c.o.m. potential at $r<  r_{c}$. Interestingly, there exists a deep ``pocket" for $^{212}$Po while a very shallow one for $^{210}$Po. This ``pocket" is of particular importance in calculating the c.o.m. wave function $\Phi(r)$.

 \begin{figure}[thb]
\includegraphics[width=0.6\textwidth]{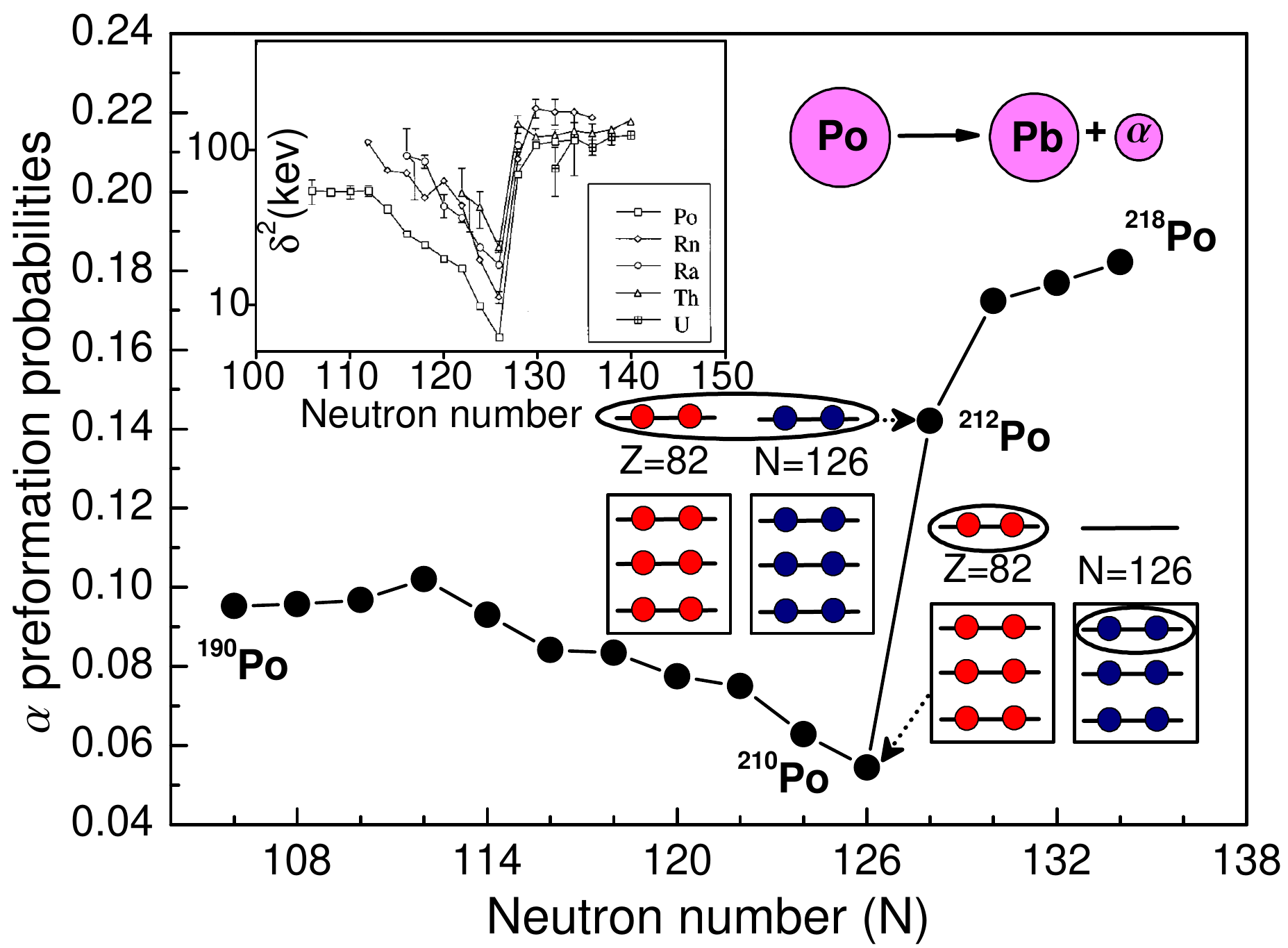}
\caption{(Color online)  Alpha-cluster preformation probability $P_\alpha$ of even-even Po isotopes by the quartetting wave function approach. For comparison, the empirical analysis of $\alpha$-decay reduced width $\delta^2$ taken from Ref.~\cite{Batchelder} are given in the insert.}
 \label{Fig:2}
 \end{figure}

The c.o.m. wave functions $\Phi(r)$ are also numerically computed for $^{210}$Po and $^{212}$Po. As shown in Fig.~\ref{Fig:1}, both wave functions exhibit an approximately linear increase up to the region of the critical radius and then decreases. However, the c.o.m. wave function of $^{212}$Po is more extended to the surface region ($r > r_c$) because of the deep ``pocket". This is different to the case of $^{210}$Po. As the $\alpha$-cluster formation probability is calculated by an integral of $\Phi(r)$ in the surface region ($r > r_c$) (see Eq.~(1)), the difference between the c.o.m. wave functions of $^{210}$Po and $^{212}$Po explains why
there is an abrupt change across the $N=126$ major shell. As seen in Fig.~\ref{Fig:1}, the deep ``pocket" of $^{212}$Po yields a bound state energy $E_{\rm tunnel}=-19.346$ MeV very close to the Fermi energy $\mu_4=-19.771$ MeV, which is also different from the case of $^{210}$Po.

 \begin{figure}[th]
\includegraphics[width=0.6\textwidth]{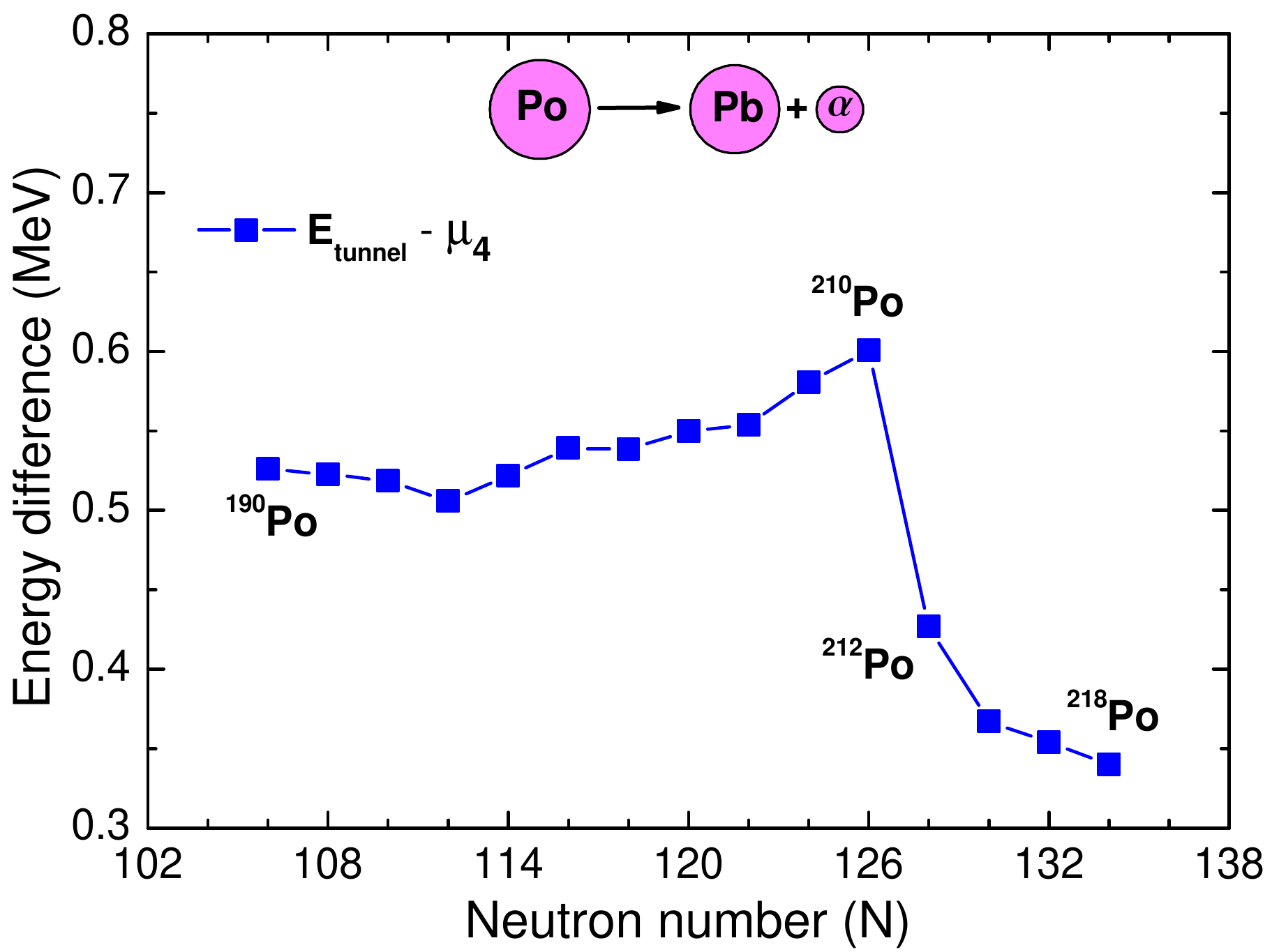}
\caption{(Color online)  The quantity $E_{\rm tunnel}-\mu_4$ as a function of the neutron number $N$ for the Po isotopes.}
 \label{Fig:3}
 \end{figure}

It is of interest to present the calculated $\alpha$-cluster formation probability $P_{\alpha}$ of all even-even polonium isotopes by using our quartetting wave function approach. A plot of $P_{\alpha}$ as a function of neutron number $N$ is given in Fig.~\ref{Fig:2}. {For comparison, in the insert the empirical values of the $\alpha$-decay reduced width $\delta^2= 2 \pi \hbar \lambda /P$ are shown, as given in Ref.~\cite{Batchelder}. Here, $\lambda$ is the measured decay constant and
$P$ penetrability factor for the $\alpha$ particle to tunnel through the Coulomb and centrifugal barriers.
Taking the same penetration probability $P$ as ${\cal T}$ calculated in our quartetting wave function approach
as introduced above, we have
the relation $P_\alpha$=$\delta^2$/(2$\pi\nu$). Therefore both quantities, the empirical $\alpha$-decay reduced width $\delta^2$ and the formation probability $P_{\alpha}$, are closely related, and the behavior of $P_{\alpha}$ in Fig.~\ref{Fig:2} is in excellent agreement with the empirical analysis \cite{Batchelder}.} As shown in Fig.~\ref{Fig:2}, the values of $P_{\alpha}$ are quite close to each other from $^{190}$Po to $^{196}$Po. This feature is called as the ``saturation effect" in previous studies \cite{Batchelder}. Then the values of $P_{\alpha}$ decrease with the increasing of neutron number until a sudden jump from $P_{\alpha}$($^{210}{\rm Po})=0.054$ to $P_{\alpha}$($^{212}{\rm Po})=0.142$ appears. This is clearly due to the effect of $N=126$ neutron shell closure. As demonstrated by the two sketches in Fig.~\ref{Fig:2}, the formation of an $\alpha$-cluster in $^{210}$Po involves two neutrons below the major closed shell. This is in contrast to the case of $^{212}$Po, in which two neutrons are on top of the lead core.
In contrast to the empirical analysis, here the formation probabilities are consistently computed in a microscopic way, which reveals very clearly the shell structure of heavy nuclei. It is also found that there exists strong correlation between the difference of $E_{\rm tunnel}-\mu_4$ and the $\alpha$-cluster formation probability, cf. Ref.~\cite{Xu2016}. A systematic
dependence of the energy difference $E_{\rm tunnel}-\mu_4$ on the neutron numbers $N$ is shown in Fig.~\ref{Fig:3} for the Po isotopes. As clearly shown by Fig.~\ref{Fig:2} and Fig.~\ref{Fig:3}, if the bound state energy $E_{\rm tunnel}$ is close to the Fermi energy $\mu_4$, a large $\alpha$ preformation probability is obtained.
\begin{figure}[th]
\includegraphics[width=0.6\textwidth]{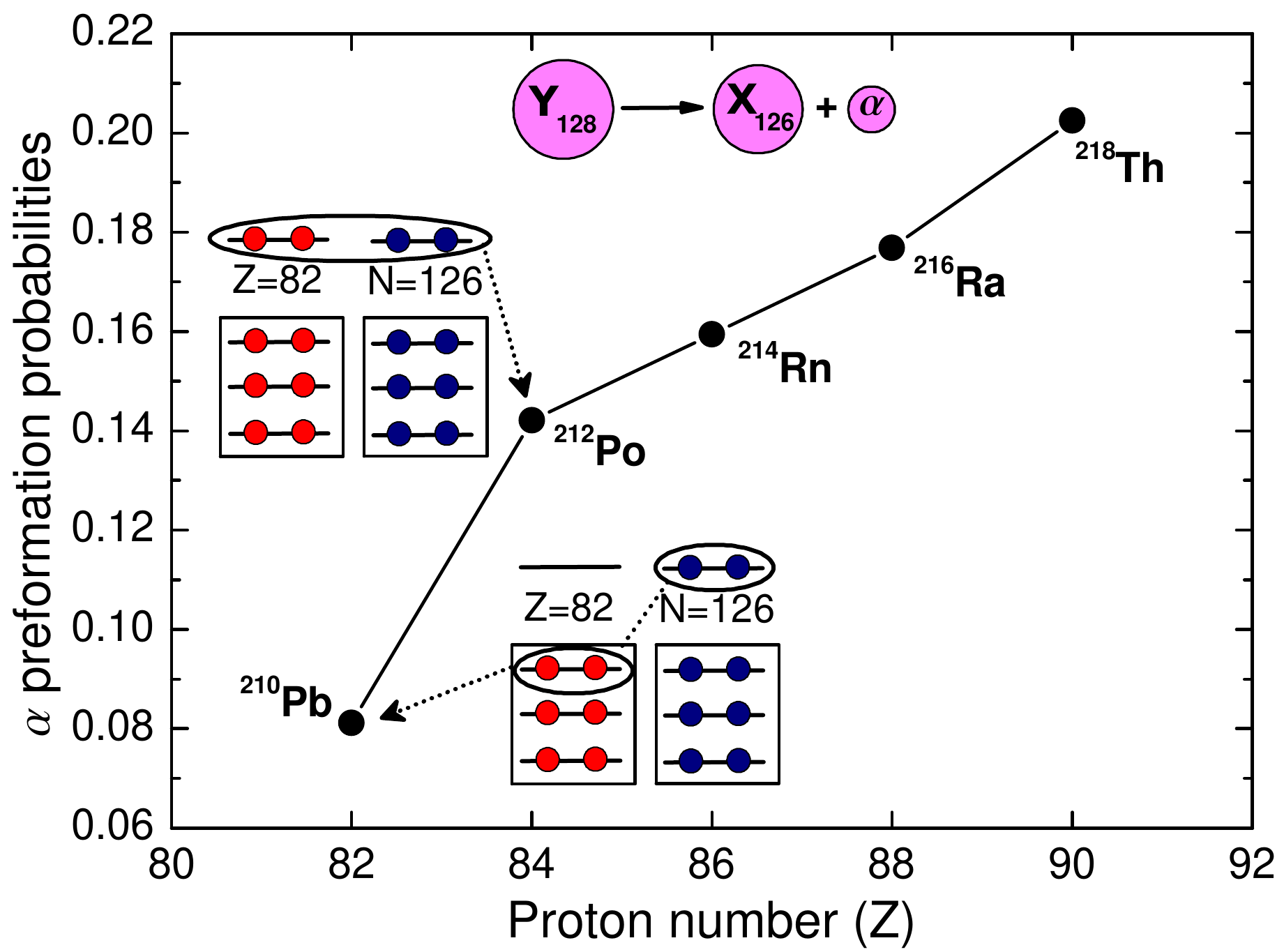}
\caption{(Color online)  Alpha-cluster preformation probability  {\bf $P_\alpha$} of even-even $N=126$ isotones by the quartetting wave function approach.}
 \label{Fig:4}
 \end{figure}

Another typical example is the $\alpha$-cluster formation probability of the even-even $N=126$ isotones (daughter nuclei). We show in Fig.~\ref{Fig:4} the  calculated $\alpha$-cluster formation probability of $\alpha$-emitters $^{210}$Pb, $^{212}$Po, $^{214}$Rn, $^{216}$Ra, and $^{218}$Th as a function of proton number $Z$. As expected, there is large increase of $\alpha$-cluster formation probability from $P_{\alpha}$($^{210}{\rm Pb})=0.081$ to $P_{\alpha}$($^{212}{\rm Po})=0.142$ because of the effect of $Z=82$ closed shell.

\begin{table*}[htb]
\caption{The $\alpha$-cluster formation probabilities of even-even superheavy nuclei by the quartetting wave function approach.}\label{Tab:1}
\begin{tabular}{|c|c|c|c|c|c|c|c|c|c|c|}
\hline \hline
Mass &$Z$ &$N$ &$Q_\alpha$ &  Half-life &$c$&$d$ & Fermi energy   &$E_{\rm tunnel}$ &
$E_{\rm tunnel}-\mu_4$ & $P_\alpha$ \\
 & & & MeV& $T_{1/2}$[s]& [MeV fm]& [MeV fm]& $\mu_4$[MeV] &[MeV] & [MeV] &     \\
\hline
294	&118 &176 &11.810  &1.4$\times 10^{-3}$  &17066.70	&4847.61 &	-16.889	 &	-16.490	 & 0.399 & 	0.110\\
292	&116 &176 &10.774  &2.4$\times 10^{-2}$  &19237.20	&5365.62 &	-17.772	 &	-17.526	 & 0.246 & 	0.197\\
290	&116 &174 &10.990  &8.0$\times 10^{-3}$  &19027.50	&5315.41 &	-17.568	 &	-17.310	 & 0.258 & 	0.191\\
288	&114 &174 &10.072  &7.5$\times 10^{-1}$  &18743.70	&5251.07 &	-18.549	 &	-18.228	 & 0.320 & 	0.156\\
286	&114 &172 &10.370  &3.5$\times 10^{-1}$  &17237.40	&4892.79 &	-18.349	 &	-17.930	 & 0.419 & 	0.104\\
270 &\textbf{110} &\textbf{160} &11.117  &2.1$\times 10^{-4}$  &17079.10	&4847.45 &	-17.547	 &-17.183	 & \textbf{0.364} &  \textbf{0.144}\\
268 &\textbf{108} &\textbf{160} &9.623   &1.4$\times 10^{0}$   &15653.10	&4516.39 &	-19.171	 &-18.677	 & \textbf{0.494} & 	\textbf{0.077}\\
264 &108 &156 &10.591  &1.1$\times 10^{-3}$  &17054.60	&4843.76 &	-18.088	 &	-17.709	 & 0.379 & 	0.140\\
260 &106 &154 &9.901   &1.2$\times 10^{-2}$  &17488.80	&4948.93 &	-18.759	 &	-18.399	 & 0.360 & 	0.152\\
\hline
\hline
\end{tabular}
\end{table*}

For {\it Superheavy nuclei}, shell effect is considered to be the determining factor for their stability. In the majority of cases, the superheavy nuclei decay via $\alpha$-decay. It is expected that the shell effect also manifests itself in the $\alpha$-cluster formation in superheavy nuclei. In Table~\ref{Tab:1}, the details of the calculated $\alpha$-cluster formation probabilities of all even-even superheavy nuclei available are presented. The parameter values $c$ and $d$ in the nucleon-nucleon interaction are carefully chosen so that both the measured decay energy $Q_{\alpha}$ and half-life $T_{1/2}$ are reproduced \cite{Audi}. Also, the experimental bound state energy $E_{\rm tunnel}=Q_{\alpha}-28.3$ MeV is reproduced, which is above the value of Fermi energy $\mu_4$. Interestingly, an {abrupt} jump of $\alpha$-cluster formation probability from $^{268}108$ ($P_{\alpha}=0.077$) to $^{270}110$ ($P_{\alpha}=0.144$) is observed. They belong to the $N=160$ isotones. This shows that $Z=108$ is a possible proton shell (or subshell) closure in superheavy mass region. More discussion of $Z=108$ shell effect can be found in Ref.~\cite{Dvorak}.
 \begin{figure}[th]
\includegraphics[width=0.7\textwidth]{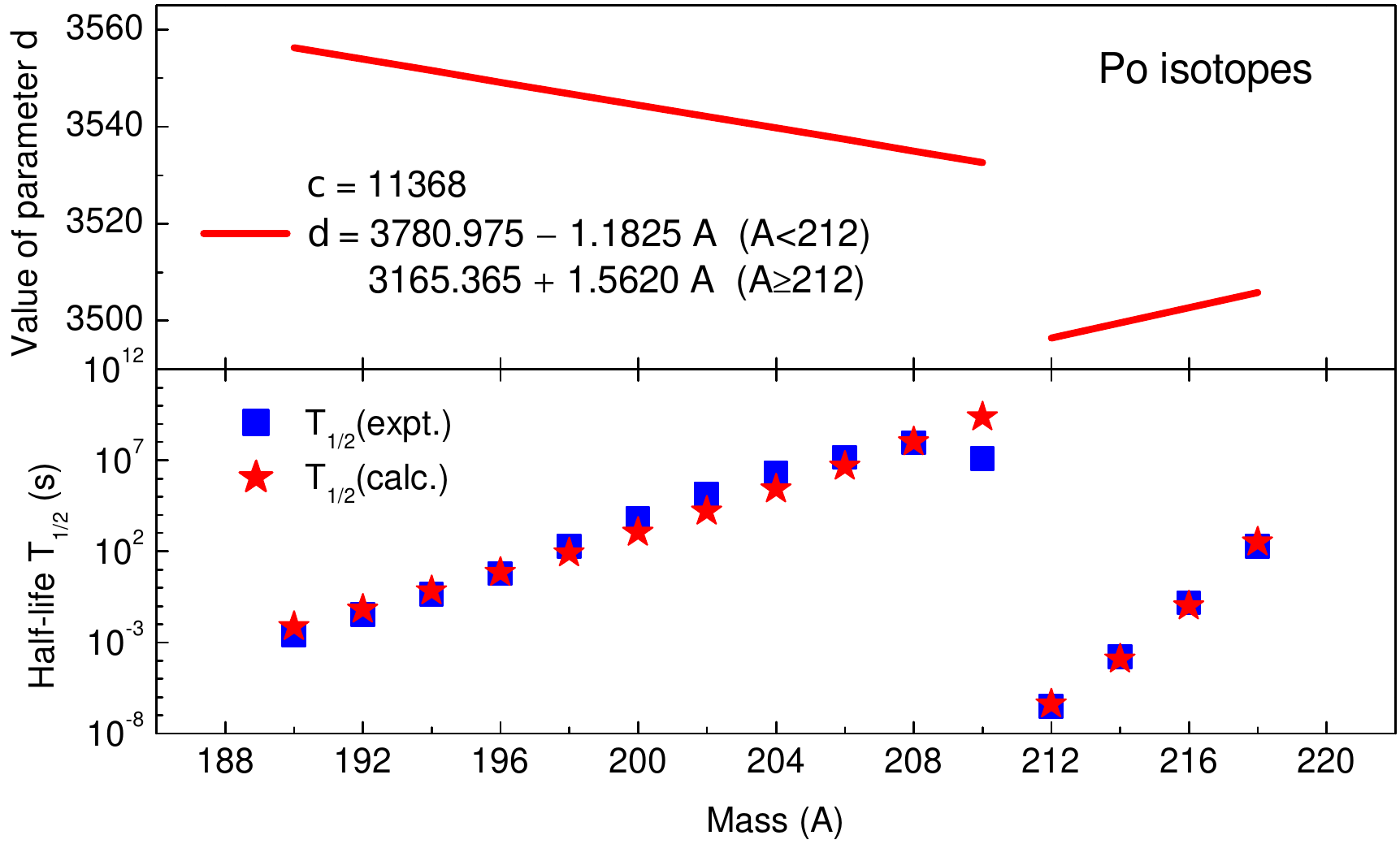}
\caption{(Color online)  The comparison of experimental and calculated half-lives for the Po isotopes by using linear mass depending parametrization of M3Y interaction strengths.}
 \label{Fig:5}
 \end{figure}

\begin{figure}[th]
\includegraphics[width=0.7\textwidth]{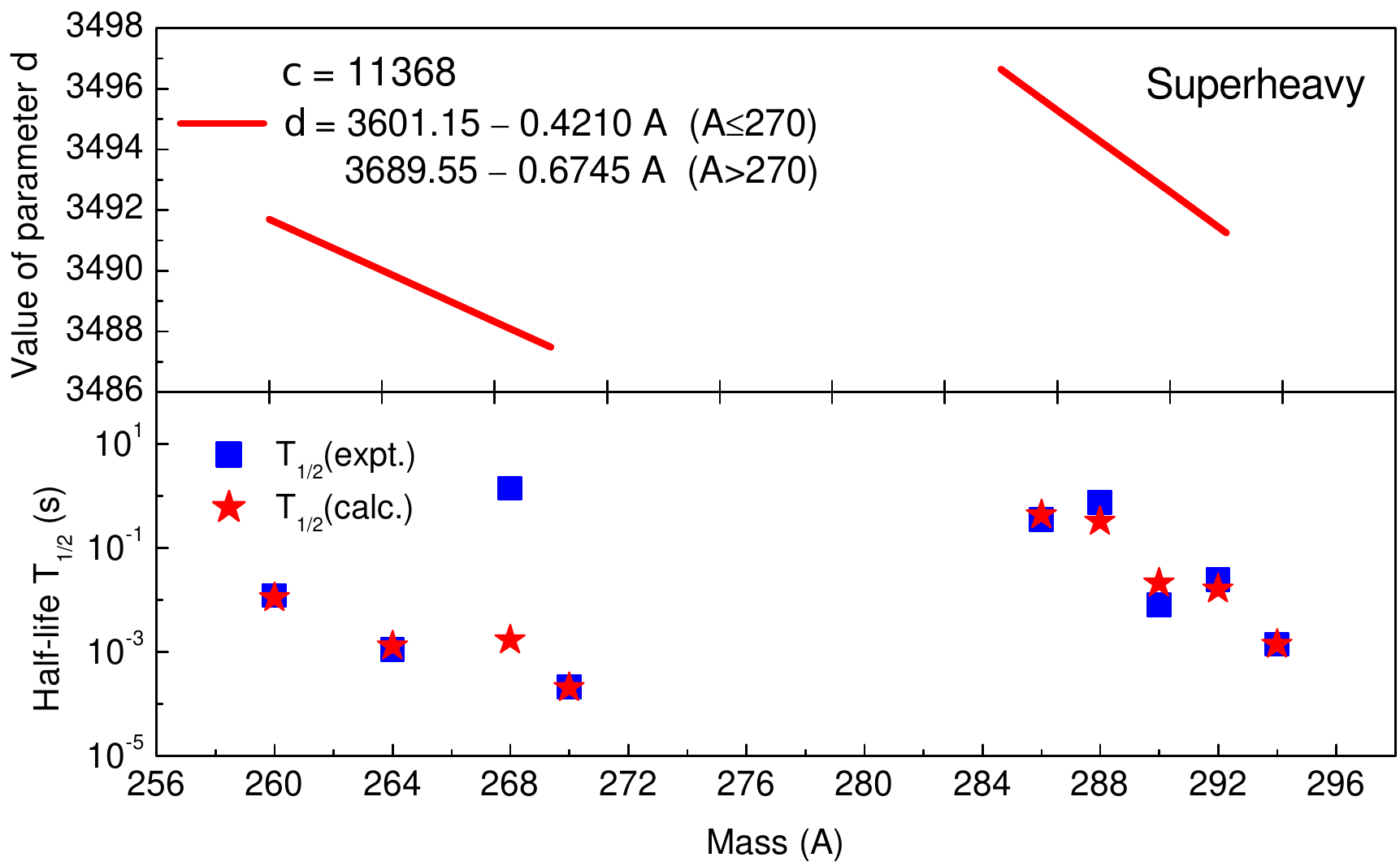}
\caption{(Color online)  The comparison of experimental and calculated half-lives for the superheavy nuclei by using linear mass depending parametrization of M3Y interaction strengths.}
\label{Fig:6}
\end{figure}

{\it Discussion on model parameters.} By adjusting the strengths {\it c} and {\it d} in the M3Y interaction for each nucleus, we find the gap in the $\alpha$-formation probability $P_{\alpha}$ which is a signature of shell closure. At present we are not explicitly performing shell structure calculations for the core nucleus. An attempt is made to construct a ``smooth" mass depending parametrization of M3Y interaction strengths. As shown in Fig.\ref{Fig:5}, the strength {\it c} of M3Y interaction is chosen as an arbitrary constant and the strength {\it d} is found to have linear mass dependence for the Po isotopes with $A<212$ and $A \ge 212$, respectively (see the upper panel). A comparison between the experimental and calculated half-lives is given in the lower panel of Fig.\ref{Fig:5}. This may be considered as predictive power of our approach where only five parameters are fitted to describe the half-lives of the entire range of Po isotopes. A similar analysis is presented for superheavies in Fig.\ref{Fig:6} where the strength {\it c} is the same constant but {\it d} has different linear mass dependence with $A \le 270$ and $A > 270$. For both the Po isotopes and superheavies, a reasonable agreement between data and theory is obtained, but deviations occur at shell closures, e.g. $N=126$ and $Z=108$. To further improve the agreement and make reliable prediction, a rigorous treatment of shell structure for the core nucleus which is missing in the present calculation based on a local-density (Thomas-Fermi) approach will be needed in future.

{\it Summary.} The physics of cluster formation in heavy and superheavy nuclei is not fully understood and
the numerical treatment is quite complex. In this work, we consider $\alpha$-cluster preformation in both the heavy and superheavy nuclei. The approach presented here to include four-nucleon correlations, in particular bound states,
is based on a first-principle approach to nuclear many-body systems. The c.o.m. motion as a collective
degree of freedom is introduced to characterize the cluster and the equation for c.o.m motion is numerically solved. An important point is that the $\alpha$-cluster can only be formed on the surface region of the core with $r>r_c$ because of the Pauli blocking effects. The interplay of the mean field and the Pauli blocking as a consequence of antisymmetrization leads to the formation of a ``pocket" in the effective c.o.m. potential. We found a deeper ``pocket" results in a more extended c.o.m wave function on the surface region and consequently a larger cluster formation probability. Systematics of $\alpha$-cluster formation probability in heavy nuclei is well reproduced by our quartetting wave function approach. We also found that the shell (or subshell) effect of $Z=108$ manifests itself in $\alpha$-cluster formation for superheavy nuclei which has been derived from data, see Table~\ref{Tab:1}.

\textit{Acknowledgments.}  This work was  discussed and prepared during the  Workshop on Nuclear Cluster Physics (WNCP2016) in Yokohama, Japan (organizer T. Yamada).
The work is supported by the National Natural Science Foundation of China (Grants No.11575082, No.11235001, No.11535004, No.11375086, and No.11120101005).

\end{document}